# Stability of Multipole-mode Solitons in Thermal Nonlinear Media

Liangwei Dong[1*], Fangwei Ye [2]

[1]*Instiue of Information Optics of Zhejiang Normal University, Jinhua, China, 321004*

[2]*Department of Physics, Centre for Nonlinear Studies,*

*and The Beijing-Hong Kong-Singapore Joint Centre*

*for Nonlinear and Complex System(Hong Kong),*

*Hong Kong Baptist University, Kowloon Tong, China*

*\*Email: donglw@zjnu.cn*

We study the stability of multipole-mode solitons in one-dimensional thermal nonlinear media. We show how the sample geometry impacts the stability of mutlipole-mode solitons and reveal that the tripole and quadrupole can be made stable in their whole domain of existence, provided that the sample width exceeds a critical value. In spite of such geometry-dependent soliton stability, we find that the maximal number of peaks in stable multipole-mode solitons in thermal media is the same as that in nonlinear materials with finite-range nonlocality.

Solitons in nonlinear media can take a variety of forms and shapes. In addition to the node-less amplitude distributions featured by fundamental solitons, solitons could also display complex spatial shapes. In one-dimensional system, these are dipole-modes which are composed of two out-of-phase fundamental solitons, or multipole-mode solitons which are composed of several fundamental solitons with alternating phases. Due to the repulsive forces between the peaks of them, multipole soliton can not exist in uniform media with local nonlinear responses, except in the form of vector solitons under suitable conditions. However, such situation changes significantly in nonlocal nonlinear media, where, in sharp contrast to the case in local media, out-of-phase bright solitons can attract each other and may even, form scalar bound states [1].

The nonlocality of the nonlinear response is a generic property of nonlinear materials. Nonlocality arises when the light-matter interaction involves mechanism such as diffusion of carriers, reorientation of molecules, heat transfer, etc [2, 3]. Recent interest in the study of nonlocal solitons has been stimulated by the experiments in nematic liquid crystals (NLCs) [4-7] as well as in thermal materials such as lead glasses [8, 9]. The nonlocal nonlinear responses of these two materials could be quite different. The nonlinear response of a NLC is characterized by a finite nonlocality degree, thus when the size of the sample is large enough relative to the spatial extent of the light, the boundary of a NLC does not affect the light dynamics or soliton properties. In contrast, in a thermal medium, the nonlinear response is always greatly determined by the details of heat diffusion at sample boundaries and thus the nonlocality degree is naturally "infinite" as the afar boundaries could significantly change the light dynamics and soliton properties [12-18], even when spatial extent of the light beam is significant small compared with that of the sample. Especially, numerical simulations [14] have demonstrated that the soliton stability may depend crucially on the sample-geometry and a rectangular sample with a proper aspect ratio may stabilize the otherwise unstable dipole modes in a square sample. Note that such strongly sample-geometry-dependent soliton properties are specific to thermal materials and are usually absent in both local and finite-range nonlocal materials.

The present work is motivated by the study of Z. Xu *et al*. in Ref. [11], where the authors have revealed that, in a nonlinear material with a finite-range nonlocality such as in a NLC, the maximal number of peaks in *stable* multipole-mode solitons is four and the modes containing more than four peaks are always unstable. As the authors in Ref. [11] deal with a material with a finite-range nonlocality, no boundary effects have been took into consideration. In contrast, in nonlocal thermal materials, boundary is always crucial, and therefore, one may naturally ask, how a thermal nonlinear material affects the stability of multipoles? Especially, one might ask whether a nonlinear response of an infinite-range nonlocality supports stable multipoles composed of more peaks than that supported by a finite-range nonlocality? It is worthy to mention that, the physical model describing light propagation in a NLC without the static bias electric field is mathematically equivalent to that in a thermal nonlinear medium, thus, boundary effects may also come into play in a NLC, as numerically analyzed and experimentally observed in [24,25]. Therefore our present study may also find potential applications in liquid crystals, although we following put our research in the context of thermal nonlocal materials.

The purposes of this paper are twofold. First, we study whether the sample-geometry-dependent soliton stability in thermal materials, which was revealed in two-dimensional settings where the controlling parameter being the sample aspect ratio, also persists in one-dimensional geometry when the controlling parameter simply being the sample width. Indeed, we find that the stability of multipole-mode solitons depends crucially on the physical dimension of the thermal material, and complete stability is achieved for tripole- and quadrupole-mode solitons when the width of the sample exceeds some crucial value. We are able to build this important finding on a rigorous linear stability analysis, which is absent in higher-dimensional settings [14] due to the requirement of huge computational resources. The second purpose of this paper is to reveal the maximum number of peaks in stable multipoles that hold in a thermal nonlinear medium. This purpose stems from the first one, as the stability characteristics of multipoles now crucially depends on sample width and thus the stability analysis for NLCs [11] does not hold in thermal nonlinear media. Surprisingly, we found that a nonlinear thermal medium possesses the same restriction on the maximum number of peaks in stable multipoles as in the case of a finite-range nonlocal medium. In another word, the maximal number of peaks in stable multipoles supported by thermal materials is also four and all modes composed of five peaks or more are unstable.

We consider the propagation of a laser beam along the $z$ axis of a thermal medium occupying the region $-L/2 \leq x \leq +L/2$ ( $L$ is the sample width), which can be described by the system of equations for the dimensionless field amplitude $q$ and the nonlinear contribution to the refractive index $n$ that is proportional to the temperature variation, given by

$$i\frac{\partial q}{\partial z} = -\frac{1}{2}\frac{\partial^2 q}{\partial x^2} - qn,$$
$$\frac{\partial^2 n}{\partial x^2} = -|q|^2, \quad (1)$$

Here the transverse and longitudinal coordinate $x, z$ are scaled to the beam width and to the diffraction length, respectively. We assume that the opposite boundaries of the thermal medium are thermostabilized, thus, $q, n|_{x=\pm L/2} = 0$.

We search for soliton solutions of Eqs. (1) in the form $q = w(x)\exp(ibz)$, where $w$ is a real function and $b$ is the propagation constant. Such solitons can be characterized by their energy flow $U = \int_{-L/2}^{L/2}|q|^2\,dx$, which is a conserved quantity of Eqs.(1). The refractive index induced by the soliton is given by $n = -\int_{-L/2}^{L/2} G(x,x')|q(x')|^2\,dx'$, where the response function $G(x,x') = L^{-1}(x+L/2)(x'-L/2)$ for $x \leq x'$ and $G(x,x') = L^{-1}(x'+L/2)(x-L/2)$ for $x \geq x'$. To elucidate the linear stability of solitons we searched for perturbed solution in the form of $q(x,z) = [w(x) + u(x)\exp(i\delta z) + v^*(x)\exp(-i\delta^* z)]\exp(ibz)$, where the perturbation modes $u$ and $v$ can grow with a complex rate $\delta$ upon propagation("*" stands for complex conjugate). Linearization of Eqs. (1) around stationary solution $w(x)$ yields,

$$\delta u = \frac{1}{2}u_{xx} - bu + un + w\int_{-L/2}^{L/2} G(x,x')w(x')[u(x') + v(x')]dx',$$
$$\delta v = -\frac{1}{2}v_{xx} + bv - vn - w\int_{-L/2}^{L/2} G(x,x')w(x')[u(x') + v(x')]dx', \quad (2)$$

Figure 1 (a) presents an example for fundamental solitons. As shown in the figure, the induced refractive index $n(x)$ is nonzero across the whole sample, which is due to the infinite range of nonlocality. Note also that the $n(x)$ produced by the dipole displays a maximum platform near $x=0$ [Fig. 1 (b)], rather a small dip encountered in a NLC [11]. Such index platform expands for lower-power dipoles [Fig. 1(c)]. The dependence of energy flow $U$ versus propagation constant $b$ is shown in Fig.1 (d). We find that fundamental and dipole solitons are stable in their whole domain of existence.

Thermal nonlinear media in principle support bound states with arbitrary number of peaks. This can be explained from their associated index profiles, as $n(x)$ always features a parabolic-like shape across the whole light regions [Fig. 2(a), Fig. 2(d) and Fig. 3(a)]. The stability analysis for tripole and quadrupole-mode solitons is summarized in Fig.2. An important result is that the stability of solitons is strongly dependent on the sample width, and instability domain shrinks quickly with the increase of the sample width [Fig. 2(c) and (f)]. For example, when $L=20$, tripole modes are stable in the region $b>b_{cr}\approx 0.7$. However, when $L$ is increased to $L=40$ the stable region expands to $b>b_{cr}\approx 0.15$ [Fig. 2 (b)]. Further, completely stable are achieved if $L>60$. Such sample-width-dependent stability characteristics are also observed for quadrupole solitons [Fig. 2(e)]. However, it should be mentioned that the instability region, defined by $b\subset[0,b_{cr}]$, for quadrupoles is wider than that for tripole. Also the instability growth rate $\delta_i$ (imaginary part of $\delta$) for quadrupole is higher than that for tripoles. Note that, as mentioned above, such geometry-dependent stability have been reported recently in two-dimensional thermal media for dipole solitons [14], where the dipole modes with *correct* orientation with respect to sample geometry are found to be stable provided that the sample is cut into a suitable geometrical aspect ratio. However, the finding in [14] is based on the propagation simulations while a linear stability analysis is absent as the latter requires huge computational resources. In contrast, here we work on a most fundamental system, thus allowing us to build the important result of geometry-dependent-stability on a rigorous linear stability analysis.

Another important finding of our stability analysis is that, although the soliton stability in thermal materials has a profound dependence on the sample-width as discussed above, the maximum number of peaks in stable multipoles is still four, which is the same as in a nonlinear medium with a finite-range nonlocality. Thus, bound states incorporating five [Fig.3 (a)] or more than five peaks are always oscillatory unstable. Such global instability holds irrespective of variation of energy flow [Fig.3 (b)] or sample width, although the specific value of $\delta_i$ quantitatively depends on both. The same restriction on the maximal number of peaks in stable multipoles for a thermal medium and a NLC might counter one's intuition, as in NLC the instability domain decreases with the increase of nonlocality degree [11], thus for a thermal medium with an infinite range of nonlocality and a strong dependence of stability on sample-geometry, one might expect that the maximal number of peaks in stable multipoles might be larger than four or at least different with that. However, our results here reveal that thermal nonlinear media actually possess the same upper threshold for stability as that in NLCs.

All results reported above are confirmed by the outcome of the direct numerical integration of Eqs. (1) with the input condition $q(x, z = 0) = w(x)[1 + \rho(x)]$, where $w(x)$ is the profile of the stationary wave and $\rho(x)$ stands for the broadband input noise with the variance $\sigma_{\text{noise}}^2 = 0.01$. As expected, the multipole-mode that belong to the stable region predicted by linear stability analysis retain their shapes over indefinitely long propagation distance, as clearly seen in Fig. 4(a) for a dipole mode, and Fig. 4(d) for a quadrupole mode. Note that, Figure 4(c) shows propagation result for another quadrupole mode at a smaller sample width (all the other parameters are the same as those in Fig. 4(d)). In contrast with that in wider sample, this quadrupole experiences oscillatory instability and destroys itself. Finally, all fifth-order solitons experience the similar instability scenario [Fig.4 (b)].

In conclusion, we studied the stability characteristics of bound states composed of several field peaks in thermal nonlinear media with an infinite-range nonlocality. We found that the stability of tripole- and quadrupole-mode soliton depends crucially on the sample width, and they can be completely stable when the sample width exceeds a critical value. We based the result of sample-width-dependent soliton stability on a rigorous linear stability analysis with collaboration by direct propagation simulations. Thermal nonlinear media cannot support stable bound modes containing more than four peaks, thus they possess the same restriction on the maximal number of peaks in sable multipoles as that in nonlinear media with finite-range of nonlocality, although their nonlocal nonlinear responses are essentially different.

This work was supported by the National Natural Science Foundation of China (Grant No. 10704067).

**Figure captions**

FIG. 1. (Color online) (a) Profile of a fundamental soliton at $b = 0.5$. Profiles of dipole solitons at (b) $b = 0.5$ and (c) $b = 5$, respectively. The thick line stands for the light field ($w$) while the light line stands for refractive index ($n$). (d) the energy flow versus propagation constant diagram for fundamental(dashed line), dipole(light line) and tripole(thick line) solutions. The two points marked by circles corresponds to the soliton solutions in (b) and (c) respectively. In all plots $L = 60$. All quantities are plotted in arbitrary dimensionless units.

FIG. 2. (Color online) (a) Profile of a tripole soliton at $b = 1$ and $L = 40$. (b) Instability growth rate for tripole solitons versus propagation constant at $L = 20, 30$ and $40$. (c) Critical value of propagation constant (above which solitons are stable) for tripole solitons versus sample width. (d) Profile of a quadrupole soliton at $b = 1$ and $L = 40$. (e) Instability growth rate for quadrupole solitons versus propagation constant at $L = 20, 40$ and $60$. (f) Critical value of propagation constant for quadrupole solitons versus sample width. All quantities are plotted in arbitrary dimensionless units.

FIG. 3 (a) Profile of a fifth-order soliton at $b = 1$. (b) Instability growth rate for the fifth-order soliton. $L = 60$. All quantities are plotted in arbitrary dimensionless units.

FIG. 4. (Color online) Propagation of a perturbed dipole soliton (a) and a fifth-order mode (b) in a sample with $L = 60$. Propagation of a perturbed quadrupole mode at (c) $L = 20$ and (d) $L = 40$. For all the plots, $b = 1$. All quantities are plotted in arbitrary dimensionless units.

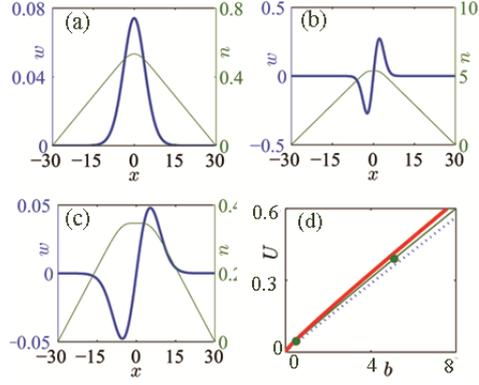

FIG. 1. (Color online) (a) Profile of a fundamental soliton at $b = 0.5$. Profiles of dipole solitons at (b) $b = 5$ and (c) $b = 0.5$, respectively. The thick line stands for the light field ($w$) while the light line stands for refractive index ($n$). (d) Energy flow versus propagation constant diagram for fundamental (dashed line), dipole (light line) and tripole(thick line) solutions. The two points marked by circles corresponds to the soliton solutions in (b) and (c) respectively. In all plots $L = 60$. All quantities are plotted in arbitrary dimensionless units.

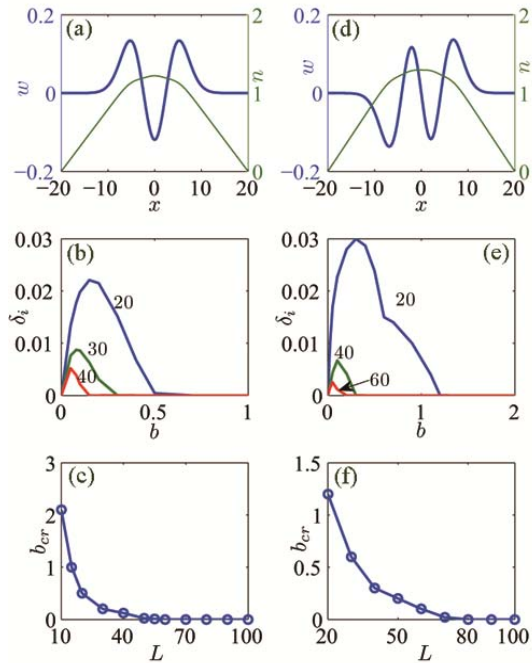

FIG. 2. (Color online) (a) Profile of a tripole soliton at $b=1$ and $L=40$. (b) Instability growth rate for tripole solitons versus propagation constant at $L=20, 30$ and $40$. (c) Critical value of propagation constant (above which solitons are stable) for tripole solitons versus sample width. (d) Profile of a quadrupole soliton at $b=1$ and $L=40$. (e) Instability growth rate for quadrupole solitons versus propagation constant at $L=20, 40$ and $60$. (f) Critical value of propagation constant for quadrupole solitons versus sample width. All quantities are plotted in arbitrary dimensionless units.

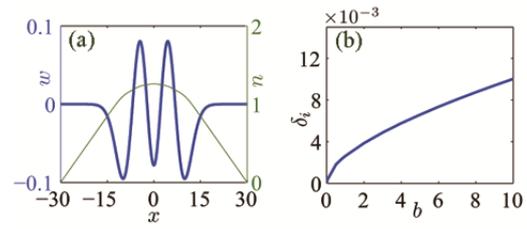

FIG. 3 (a) Profile of a fifth-order soliton at $b=1$. (b) Instability growth rate for the fifth-order soliton. $L=60$. All quantities are plotted in arbitrary dimensionless units.

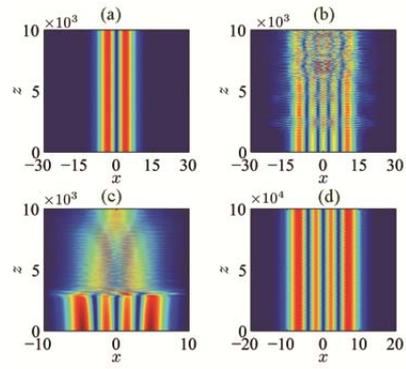

FIG. 4.  (Color online) Propagation of a perturbed dipole soliton (a) and a fifth-order mode (b) in a sample with $L=60$.  Propagation of a perturbed quadrupole mode at (c) $L=20$ and (d) $L=40$. For all the plots, $b=1$. All quantities are plotted in arbitrary dimensionless units.